\documentclass[
prc,aps,amsmath,floatfix,superscriptaddress,showpacs
]{revtex4}
\usepackage{bm}
\usepackage{amsmath}
\usepackage{epsfig}
\begin{document}
\title{In-medium modification of the isovector pion-nucleon amplitude}
\author{G. Chanfray}
\affiliation{IPN Lyon, IN2P3-CNRS et UCB
 Lyon I, F-69622 Villeurbanne Cedex} 
\author{M. Ericson}
\affiliation{IPN Lyon, IN2P3-CNRS et UCB
 Lyon I, F-69622 Villeurbanne Cedex} 
\affiliation{Theory division, CERN, CH-12111 Geneva} 
\author{M. Oertel} 
\affiliation{IPN Lyon, IN2P3-CNRS et UCB
 Lyon I, F-69622 Villeurbanne Cedex}

\begin{abstract}
We study the in-medium modification of the isovector $\pi N$ amplitude 
using a non-linear representation of the sigma model but keeping the
scalar degree of freedom. We check that our  result does not depend on
the representation. We discuss the connection with other approaches based on 
chiral perturbation theory. 
\end{abstract}
\maketitle

 \section{Introduction}
The experimental program on deeply bound pionic states has revived 
the interest on the interaction of pions with nuclei. Recently the 
attention has focused on the charge exchange scattering length
$b_1$~\cite{G02, KY01, W01, KW01,F02, KKW02, Y02}. The fit to these data~\cite{G02} 
suggests a large enhancement of $b_1$ in nuclei, as 
anticipated in Ref.~\cite{W01}. It is the purpose of this letter 
to investigate this question within a non-linear representation of the
sigma model keeping the scalar degree of freedom. The advantage of 
this model is that it allows an explicit evaluation of the contribution of the 
nuclear pion gas in a way which satisfies the constraints imposed by chiral 
invariance.  At the same time it  keeps the extra degree of freedom 
associated with the scalar meson so as not to exhaust the physics 
only with pions.  The quantity that we investigate is the charge exchange 
amplitude of a quasi-pion ({\it i.e.}, on mass shell in the medium) on a nucleon 
embedded in infinite nuclear matter. We will derive the in-medium 
correction to first order in the nucleon density. Our 
emphasis is not so much on a comparison with data but rather
on the derivation itself of the in-medium value for $b_1$   and on
some important questions of principle arising in comparison with other theoretical
works. 
 
 \section{The s-wave pion self-energy}
\begin{figure}[h]
\begin{center}
\epsfig{file=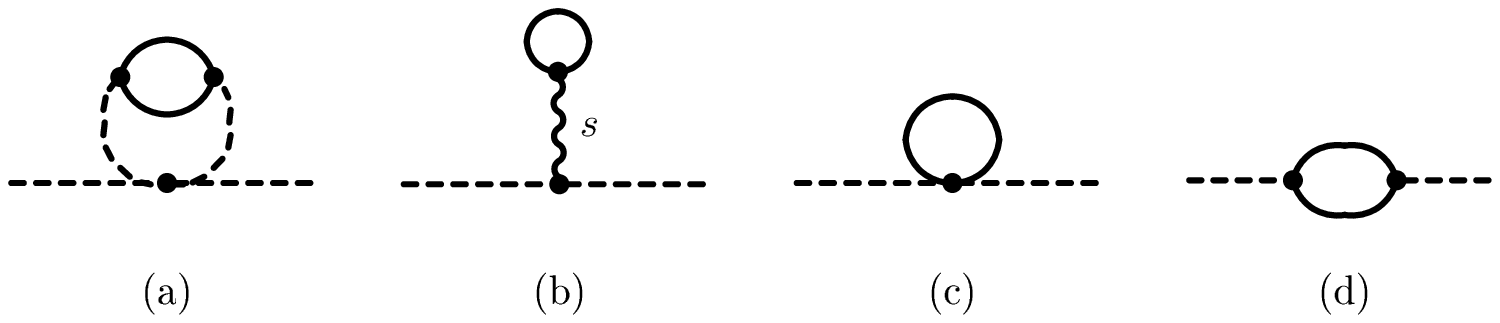,width=12.0cm,angle=0}
\end{center}
  \caption{The various contributions to the s-wave pion self-energy. More explanations
  are given in the text.}
\label{piself}      
\end{figure}

Our starting point is based on the linear sigma model but reformulated as in 
Ref.~\cite{CEG02}. The original sigma field $\sigma$ and pion fields $\vec\pi$ are
eliminated in favor of a chiral invariant scalar field, $S=f_\pi+s$, and a 
new pion field
$\vec\phi$ according to: 
\begin{equation}
\sigma\,+\,i\,\vec\tau\cdot\vec\pi=S\,U\equiv(f_\pi\,+\,s)\,
\exp\left[i{\frac{\vec\tau\cdot\vec\phi}{f_\pi}}\,G\left({\frac{\phi^2}{f^2_\pi}}\right)
\right]~.
\end{equation}
As proposed in Ref.~\cite{CEG02}, the fluctuating scalar field $s$, which
physically describes the fluctuations of the radius of the chiral
circle around its vacuum value $f_\pi$, can be identified with the
sigma meson of relativistic theories of the Walecka type. The
function $G(X^2)=1\,+\,\alpha\,X^2\,+...\,$ selects the particular
realization of the model. Changing \( G \) amounts to a redefinition
of the pion field, which should not affect physics.  In the
following we keep \( \alpha \) arbitrary and check that the final
results for physical observables do not depend on this parameter. The
Lagrangian as given by eqs.(19-30) of Ref.~\cite{CEG02} writes (notice that
the notations $\Theta, \theta$ for the chiral invariant scalar field
are now replaced by $S, s$):

\begin{eqnarray}
{\mathcal{L}} & = & \left( f_{\pi }+s \right) ^{2}\, Tr\partial ^{\mu
}U\partial _{\mu }U^{\dagger }\, +\, {1\over 2}\partial ^{\mu }s \partial
_{\mu }s \, -\, {m^{2}_{\sigma }-m^{2}_{\pi }\over 8f^{2}_{\pi }}\left(
s ^{2}+2f_{\pi }s \, +\, {2f^{2}_{\pi }m^{2}_{\pi }\over m^{2}_{\sigma
}-m^{2}_{\pi }}\right) ^{2}\nonumber \\
 &  & +i\bar{N}\gamma ^{\mu }\partial _{\mu }N\, -\, M_{N}\left( 1+{s \over
 f_{\pi }}\right) \bar{N}N\nonumber \\
 &  & +\bar{N}\gamma _{\mu }{\mathcal{V}}^{\mu }_{c}N\, +\left( 1\,
 -(1-g_{A})\left( 1+{s \over f_{\pi }}\right) ^{2}\right) \, \bar{N}\gamma
 _{\mu }\gamma ^{5}{\mathcal{A}}^{\mu }_{c}N\nonumber \\
 &  & +i{1-g_{A}\over 2f_{\pi }}\left( 1+{s \over f_{\pi }}\right)
 \bar{N}\gamma ^{\mu }N\, \partial _{\mu }s +{\mathcal{L}}_{\chi
 SB},\label{Eq21} 
\end{eqnarray}
where we have defined (with $\xi^2=U$): 
\begin{equation}
\label{Eq22}
{\mathcal{V}}^{\mu }_{c}={i\over 2}\left( \xi \partial _{\mu }\xi ^{\dagger
}+\xi ^{\dagger }\partial _{\mu }\xi \right) \, \qquad {\mathcal{A}}^{\mu
}_{c}={i\over 2}\left( \xi \partial _{\mu }\xi ^{\dagger }-\xi ^{\dagger
}\partial _{\mu }\xi \right) .
\end{equation}
This field transformation has the advantages of the non-linear 
realization but the radius of the chiral circle is not frozen and it  
keeps the degree of freedom associated with the scalar 
meson. The new pion field couples derivatively (to both the nucleon and 
the scalar 
field) in such a way that the low-energy theorems are satisfied without 
need of cancellations.
 As for the symmetry breaking piece
we have the following expression in terms of the new variables: 
\begin{equation}
\label{Eq23}
{\mathcal{L}}_{\chi SB}=f_{\pi }m^{2}_{\pi }\, \big (f_{\pi }+s \big
)\,\cos \left({\phi\over f_\pi} G\left( \frac{\phi^2 }{f_{\pi }^2}\right) \right)~.
\end{equation}
In order to reproduce the experimental value of the 
isoscalar scattering length it is necessary, as in chiral perturbation 
theory, 
to introduce additional chiral invariant pieces contributing to order ${\cal O}(Q^2)$: 
\begin{equation}
\Delta{\cal L}^{(2)}_{\pi N}=c'_3\,\bar N(u\cdot u)N\,+\,c'_2\,
\bar N(v\cdot u)^2 N~,\label{EXTRA}
\end{equation}
where $u_\mu=i\xi^\dagger\partial_\mu U\xi^\dagger$ and $v_\mu$ is
the four-velocity of the nucleon. These $c'_2$ and $c'_3$ parameters have to be
empirically determined. Finally the introduction of a form  factor at
the p-wave $\pi NN$ vertex allows the contact with the standard 
pion-nucleon and pion-nucleus phenomenology. In this model the pion-nucleon sigma 
term, $\sigma_N$, receives two contributions, one from the scalar field of the 
nucleon and one from its pion cloud~:
\begin{equation}
\sigma_N=\Sigma_N^{(s)}\,+\,\Sigma_N^{(\pi)}~.
\end{equation}
The scalar contribution is given by:
\begin{equation}
\Sigma_N^{(s)}= g_{sN}\,f_\pi\,{m^2_\pi\over m^2_\sigma}=
M_N\,{m^2_\pi\over m^2_\sigma}~,
\end{equation}
where $g_{sN}=M_N/f_\pi$ is the scalar meson-nucleon coupling constant.
In the nuclear medium  the relative amount of 
restoration  is linked to the mean scalar field $ \langle s\rangle$. To leading order
in density it is given by~: 
\begin{equation}
{\Sigma_N^{(s)}\,\rho\over f^2_\pi\,m^2_\pi}=-{\langle s\rangle\over f_\pi}~.
\end{equation}
The pion cloud contribution  (with its important $\Delta$ piece) 
has been explicitly calculated  with a 
model form factor. The resulting value is $\simeq$ 25  MeV \cite{JCT92,BMG92,LTTW00}, 
about half of the total value. 
In order to link it to the chiral perturbation result we expand 
it in orders of the pion mass:
\begin{equation}
\Sigma_N^{(\pi)}=-4\,c'_1\,m^2_\pi\,+\,(\Sigma_N^{(\pi)})^{(LNAC)}+...~.
\end{equation}
The coefficient $c'_1$ depends on the model ({\it i.e.}, on the form
factor). It is related to the coefficient $c_1$ of chiral perturbation
theory by $c'_1=c_1-M_N/4 m^2_\sigma$. The leading non analytical
contribution, $(\Sigma_N^{(\pi)})$, is instead given by chiral
symmetry alone:
\begin{equation}
(\Sigma_N^{(\pi)})^{(LNAC)}=- {9\,g^2_A\over 64\,\pi\,f^2_\pi}\,m^3_\pi~.
\end{equation}
Its numerical value is $-22$~MeV, to be compared with the
value of the full sigma term which is $\sigma_N~\simeq +~45-50$~MeV. We stress 
that such a model pion cloud calculation has
been successfully used to extrapolate lattice data into the low quark
mass, {\it i.e.}, to the low pion mass sector \cite{LTTW00}.
Introducing the scalar density of nuclear pions, $\langle\phi^2\rangle$, the
relative amount of restoration from the pion cloud is given to leading order in
density by:
\begin{equation}
{\Sigma_N^{(\pi)}\,\rho\over f^2_\pi\,m^2_\pi}=
{\langle \phi^2\rangle\over 2\, f^2_\pi}~.
\end{equation}
We now turn to the s-wave isoscalar pion self-energy for pions of zero 
three-momentum. To linear order in density it is simply given by the
vacuum $\pi N$ $T$-matrix multiplied by the density. The various
contributions are depicted in Fig.~\ref{piself}. The pion loop
contribution (fig. 1a), including vertex corrections not shown in
Fig.~\ref{piself}a, is derived in Refs.~\cite{CEG02,CEW96,CD99}.  It
involves the pionic piece of the sigma term. As for the scalar
exchange contribution (fig. 1b) it is related to the scalar piece of
the sigma term. Fig. 1c is immediately obtained from the Lagrangian
given in eq.~(\ref{EXTRA}) and fig. 1d is the Born term contribution
with pseudo-vector coupling. The final result to leading order in
density reads:
\begin{equation}
\Pi(\omega)=\rho\left\{ -\left({\sigma_N\over f^2_\pi}\,-\,{4\over 3}\, 
{\Sigma_N^{(\pi)}\over f^2_\pi}\right)\,+\, 
\omega^2\,\left({g^2_A\over 4\,M_N\,f^2_\pi}\,-\,2\,{c'_2\,+\,c'_3\,-
{\Sigma_N^{(s)}\over m^2_\pi}\over f^2_\pi}\right)\,-\,
{2\,\beta\over 3}\,(\omega^2\,-\,m^2_\pi)\, 
{\Sigma_N^{(\pi)}\over f^2_\pi\,m^2_\pi}\right\}
\label{PIMODEL}
\end{equation}
Note that the pion self-energy is not directly an observable and it 
depends on
the representation through the $\beta$ factor:
\begin{equation}
\beta=1\,+\,10\,\left(\alpha\,-\,{1\over 6}\right)~.
\end{equation}

\section{In-medium modification of the isovector pion-nucleon amplitude}
The energy dependence of the pion self-energy implies a wave-function
renormalization of the in-medium pion, as noticed in several works
\cite{CD99,CEK94,TW95,CDE98,WOM02}.  The particular importance of the
energy dependence for the interpretation of the deeply bound pionic
states has recently been emphasized in Ref.~\cite{KKW02}. The pion
propagator has the form $D=(\omega^2 - m^2_\pi - \Pi(\omega))^{-1} =
Z\,(\omega^2 - m^{*2}_\pi )^{-1}$. From the expression for the self-energy, 
eq. (\ref{PIMODEL}), we obtain for the residue~:
\begin{equation}
Z=1\,+\,\left({g^2_A\over 4\,M_N\,f^2_\pi}\,-\,2\,{c'_2\,+\,c'_3\,-
{\Sigma_N^{(s)}\over m^2_\pi}\over f^2_\pi}\,-\,{2\,\beta\over 3}\,
{\Sigma_N^{(\pi)}\over f^2_\pi\,m^2_\pi}\right)\,\rho\label{ZMODEL}
\end{equation}
The value of the isoscalar scattering length, $b_0$, provides a
relation between the different parameters entering this expression.
For the nucleon it is given by:
\begin{equation}
4 \pi\,\left(1\,+\,{m_\pi\over M_N}\right)\,b_0=
-{\sigma_N\over f^2_\pi}\,+\,{4\over 3}\, {\Sigma_N^{(\pi)}\over f^2_\pi}
\,+\,\left({g^2_A\over 4\,M_N\,f^2_\pi}\,-\,2\,{c'_2\,+\,c'_3\,-
{\Sigma_N^{(s)}\over m^2_\pi}\over f^2_\pi}\right)\, m^2_\pi \simeq 0~.
\end{equation}
According to recent accurate data \cite{S01}, this value is compatible
with zero. This  near-vanishing of $b_0$ (believed to be 
fortuitous) translates 
into the
following expression for the residue:
\begin{equation}
Z\simeq 1\,+\,{\sigma_N\,\rho\over f^2_\pi\,m^2_\pi}
\,-\,{4\over 3}\,{\Sigma_N^{(\pi)}\,\rho\over f^2_\pi\,m^2_\pi}
\,-\,{2\,\beta\over 3}\,{\Sigma_N^{(\pi)}\,\rho\over
f^2_\pi\,m^2_\pi}~.\label{RESMODEL}
\end{equation}

\begin{figure}[h]
\begin{center}
\epsfig{file=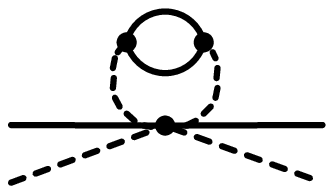,width=3.0cm,angle=0}
\end{center}
  \caption{Pion loop correction to the isovector pion-nucleon amplitude.}
\label{piwt}      
\end{figure}

The residue provides a first source for the renormalization of the
isovector scattering length $b_1$. Notice that if the pion loop
correction (the last two terms in eq. (\ref{RESMODEL})) is ignored one
recovers the original proposal of Ref.~\cite{W01}.  However this
description would be incomplete since, as previously discussed, the
pion cloud piece of the sigma term is certainly not negligible. 
In addition a description of the
in-medium modification with the only influence of the residue is
clearly unsatisfactory since the result then depends on the
representation. Indeed there is another source of renormalization,
which to our knowledge has previously been ignored.  It arises from
the pion loop correction of the isovector pion-nucleon amplitude,
inherent to the non-linear realization. The isovector piece
(Weinberg-Tomozawa)  of the chiral Lagrangian is given by:
\begin{equation}
{\cal L}_{WT}={i\over 2}\,\bar N\,\gamma^\mu \,
\left(\xi\partial_\mu\xi^\dagger +
\xi^\dagger\partial_\mu\xi\right)\, N
\simeq - {1\over 4\,f^2_\pi}\,\bar N\,\gamma^\mu \,
\left(1\,+\,2\,\left(\alpha\,-\,{1\over 24}\right)\,{\phi^2\over f^2_\pi}\right)\,
\vec\tau\cdot(\vec\phi\times\partial_\mu\vec\phi)\,N~,
\end{equation}
where the second expression arises  from an expansion to fourth order in the
pion field.  After the appropriate isospin averaging,
one obtains an effective  Weinberg-Tomozawa Lagrangian which is  modified in the medium 
by one-pion loop correction (see fig.~\ref{piwt}):
\begin{eqnarray}
{\cal L}^{eff}_{WT} &=&- {1\over 4\,f^2_\pi}\,\left(1\,+\,
{10\over 3}\,\left(\alpha\,-\,{1\over
24}\right)\,{\langle\phi^2\rangle\over f^2_\pi}\right)\,
\bar N\,\gamma^\mu \,\vec\tau\cdot(\vec\phi\times\partial_\mu\vec\phi)\,N
\nonumber\\
&=&- {1\over 4\,f^2_\pi}\,\left(1\,+\,
{20\over 3}\,\left(\alpha\,-\,{1\over
24}\right)\,{\Sigma_N^{(\pi)}\,\rho\over f^2_\pi\,m^2_\pi}\right)\,
\bar N\,\gamma^\mu \,\vec\tau\cdot(\vec\phi\times\partial_\mu\vec\phi)\,N~.
\end{eqnarray}

The isospin-antisymmetric amplitude is proportional to the pion energy 
$q_0$. In the vacuum $q_0=m_\pi$, and
$b_1=-m_\pi/(4 \pi f^2_\pi(1+m_\pi/M_N))$.
For the quasi-particle in the medium the energy is the pion 
effective mass $m_\pi^*$. Since we take $b_0$ to be zero, to first order 
in 
the density the effective and bare pion
masses are equal. According to the previous discussion the in-medium 
renormalization of $b_1$ is then~:
\begin{equation}
\frac{b_1^*}{b_1} = 
Z\,\left(1\,+\, 
{20\over 3}\,\left(\alpha\,-\,{1\over
24}\right)\,{\Sigma_N^{(\pi)}\,\rho\over f^2_\pi\,m^2_\pi}\right)~,
\end{equation}
which gives~:
\begin{equation}
{b_1^*\over b_1}=
1\,+\,{\sigma_N\,\rho\over f^2_\pi\,m^2_\pi}
\,-\,{7\over 6}\,{\Sigma_N^{(\pi)}\,\rho\over f^2_\pi\,m^2_\pi}~.
\end{equation}
Notice that once the two influences are taken into account the
representation dependence has disappeared, an important consistency
check of our result. Coming to a numerical evaluation, the parameters
of the model, in particular the sigma mass, can be fixed by a
simultaneous fit to the saturation curve and to pion-pion phase shifts
\cite{C02,CEO02}. A typical value is $m_\sigma=800$~MeV, yielding
$\Sigma_N^{(s)}\simeq 28$~MeV. Taking also $\Sigma_N^{(\pi)}\simeq
25$~MeV, we have~:
\begin{equation}
{\Sigma_N^{(s)}\,\rho\over f^2_\pi\,m^2_\pi}\simeq 0.21\,{\rho\over\rho_0},\qquad \qquad
\frac{\Sigma_N^{(\pi)}\,\rho}{f^2_\pi\,m^2_\pi}\simeq 0.18\,{\rho\over\rho_0}~.
\end{equation}
This leads to the estimate:
\begin{equation}
{b_1^*\over b_1}\,\simeq \,1\,+\,0.18\,{\rho\over\rho_0}~.
\end{equation}
Even if the coefficient $0.18$ may be slightly changed if the relative
weight between the pion cloud and the scalar contribution to the sigma
term is changed, the enhancement is more moderate than the original
proposal \cite{W01} which ignores the pion loop correction (in that case the
coefficient is of the order of $0.35-0.40$ depending on the precise
value of the sigma term). In our case this pion loop correction partly cancels the
effect of the pion-nucleon sigma term. 

\section{Comparison with a chiral perturbation approach}
The pion-self-energy has been calculated using a chiral perturbation
approach. To leading order in density, it reads \cite{PJM01, KKW02}:
\begin{equation}
\Pi(\omega)=\rho\,\left\{-\left({\sigma_N\over f^2_\pi}\,-\,{4\over 3}\, 
{(\Sigma_N^{(\pi)})^{(LNAC)}\over f^2_\pi}\right)\,+\, 
\omega^2\,\left({g^2_A\over 4\,M_N\,f^2_\pi}\,-\,2\,{c_2\,+\,c_3
\over f^2_\pi}\right)\,+\,
{\zeta\over 3}\,(\omega^2\,-\,m^2_\pi)\, 
{(\Sigma_N^{(\pi)})^{(LNAC)}\over f^2_\pi\,m^2_\pi}\right\}\label{PICHIPT}~.
\end{equation}
One can make a one-to-one correspondence between this CHIPT result and
the result obtained in our  model (eq.~\ref{PIMODEL}). The
first term ($\sigma_N$) and the third term (Born term) are identical
in both expressions of the self-energy. The fourth terms in both expressions are also
equivalent since the comparison between the two just defines the
quantity $c_2 + c_3$ in the particular model  that we use. As
for the representation dependent coefficient $\zeta$  which appears in the
last term of eq. \ref{PICHIPT}, it has to be be identified with $-2\,\beta$. 
However there is an essential  difference both at the conceptual and quantitative
levels. The second term and the fifth one (representation dependent term)
involve the LNAC piece of the pion loop, at variance with the model
calculation where the full pion loop appears. We will return later to this
difference and we study for the moment the consequences for the $b_1$
parameter.  Following the same steps as in section III, with the
constraint that the isoscalar scattering length vanishes, we obtain
for the residue:
\begin{equation}
Z= 1\,+\,{\sigma_N\,\rho\over f^2_\pi\,m^2_\pi}
\,-\,{4\over 3}\,{(\Sigma_N^{(\pi)})^{(LNAC)}\,\rho\over f^2_\pi\,m^2_\pi}
\,-\,{2\,\beta\over 3}\,{(\Sigma_N^{(\pi)})^{(LNAC)}\,\rho\over
f^2_\pi\,m^2_\pi}~.
\label{RESchipt}
\end{equation}
 Taking, as in Ref.~\cite{KKW02}, $\zeta=0$ ({\it i.e.} $\beta=0$), the influence of the
 residue on the $b_1$ parameter becomes~:
\begin{equation}
\left({b_1^*\over b_1}\right)^{(CHIPT1)}=Z=1\,+\,{\sigma_N\,\rho\over f^2_\pi\,m^2_\pi}
\,-\,{4\over 3}\,{(\Sigma_N^{(\pi)})^{(LNAC)}\,\rho\over f^2_\pi\,m^2_\pi}
\,\simeq \,1\,+\,0.53\,{\rho\over\rho_0}~,
\end{equation}
which yields a larger enhancement than our model calculation. However
this result is not representation independent. The pion-loop
correction in the Weinberg-Tomozawa amplitude which is not taken into
account in the CHIPT calculation should be added. If this is done in a
consistent way one should retain only the LNAC piece of the pion loop
in order to reach the representation independent following result:
\begin{equation}
\left({b_1^*\over b_1}\right)^{(CHIPT2)}=1\,+\,{\sigma_N\,\rho\over f^2_\pi\,m^2_\pi}
\,-\,{7\over 6}\,{(\Sigma_N^{(\pi)})^{(LNAC)}\,\rho\over f^2_\pi\,m^2_\pi}
\,\simeq \,1\,+\,0.51\,{\rho\over\rho_0}~.
\end{equation}
This is numerically very close to what is obtained from the prescription used in
Ref.~\cite{KKW02} but still difficult to reconcile with our model result.
The crucial difference is the fact that, in our model, the pion loop
always appears as a whole and the LNAC can never appear as an isolated
piece. In other words the model involves the combination
$-4\,c'_1\,m^2_\pi\,+\,(\Sigma_N^{(\pi)})^{(LNAC)}$ rather than
$(\Sigma_N^{(\pi)})^{(LNAC)}$ alone.  The two quantities are of
comparable magnitude but of opposite signs.  The difference between the 
two approaches is
associated with different
off-mass shell behavior of the $\pi N$ amplitude. As a further test of our
method, in the PCAC representation in which the relation
$\partial_\mu\vec{\cal A}^\mu=-f_\pi m^2_\pi \vec\phi$ holds, the soft
pion amplitude (obtained by taking $\omega=0$) should be proportional
to $\sigma_N$. In order to obtain the PCAC representation it is
sufficient to send the sigma mass to infinity (the sigma term then has
only the pionic contribution) and to take $\beta=1$ ({\it i.e.},
$\alpha=1/6$ ). In that case one recovers at the soft pion point
the venerable low-energy theorem
since:
\begin{equation}
\bigg({\cal M}(0,\, 0,\,0,\,0)\bigg)^{Model}_{PCAC}=-\,{\sigma_N\over f^2_\pi}
\,+\,2\,
{\Sigma_N^{(\pi)}\over f^2_\pi}={\sigma_N\over f^2_\pi}~.
\end{equation}
Instead with the
prescription of eq.~(\ref{PICHIPT}) for the CHIPT off-shell amplitude, one gets:
\begin{equation}
\bigg({\cal M}(0,\, 0,\,0,\,0)\bigg)^{CHIPT}_{PCAC}=-\,{\sigma_N\over f^2_\pi}
\,+\,2\,
{\Sigma_N^{(\pi)}\over f^2_\pi}^{(LNAC)}~.
\end{equation}
The two terms  on the r.h.s. being negative  cannot cooperate to 
reproduce the positive
sigma term, in conflict with the soft pion theorem.

\section{Remarks and Conclusion}
We have studied the in-medium renormalization of the isovector
amplitude in the linear sigma model. We use it in a non-linearized
representation in which the radius of the chiral circle is not frozen,
which preserves the sigma degree of freedom.  In this approach the
nucleon sigma commutator is built of a scalar meson exchange with the
condensate and a two-pion exchange one, the pion cloud
contribution. We find, as in our previous works on other in-medium
quantities, that the independence on the representation is achieved
through a combination of two effects : i) the residue of the pion
propagator arising from the energy dependence of the pion self-energy,
ii) the influence of the pion loops on the Weinberg-Tomozawa amplitude
(which must also apply to the chiral perturbation approach of
Refs.~\cite{KKW02,PJM01} so as to make it independent on the
representation).  Our description results in a more moderate
renormalization of $b_1$: at normal nuclear density, an enhancement by
18\%, which is less than the prediction of Ref.~\cite{W01} or the one
which can be derived from the results of Ref.\cite{KKW02, PJM01}. The
difference between our model and the CHIPT approaches lies in the role
of the pion loops. Our treatment for these is identical to the one
employed in Ref.~\cite{CEW96} for the pion gas in a heat bath. The
results of Ref.~\cite{CEW96}, when taken in the chiral limit,
reproduce the chiral perturbation expansion up to second order. The
only difference between the present calculation and the description of
the pion gas in a heat bath amounts to the (physically natural)
replacement of the scalar density of thermally excited pions by that
of the nuclear pions, a positive quantity. On the contrary, in the
chiral approach of Refs.~\cite{KKW02, PJM01} only the leading
non-analytical term appears in the self-energy (beyond what is
implicitly contained in the nucleon sigma commutator). Since
$(\Sigma_N^{(\pi)})^{(LNAC)}$ is negative and comparatively large, the
numerical consequences which follow explain the difference in the
renormalization factor.  It will be interesting to exactly clarify
why, in the chiral approach, $(\Sigma_N^{(\pi)})^{(LNAC)}$, which is in
fact a piece of the pion density, dissociates from the rest, while the
whole pion density (with its simple physical interpretation) naturally
enters in our approach. The nucleon size, which is crucial in any model
evaluation of the nuclear pion loop, introduces a scale which does not
seem to be fully incorporated in this CHIPT approach. The model
calculations of the pion loop which explicitly takes into account the
nucleon size ({\it i.e.}, in practice the form factor) have proven
their relevance and usefulness. For instance they give an excellent
fit to the quark mass dependence of the lattice results allowing a
convincing extrapolation to the physical mass region \cite{LTTW00}.
 
The actual fits of recent data on deeply bound pionic states suggest a
larger enhancement of $b_1$ than our prediction \cite{KY01,G02,Y02}.  In
this context we want to point out that the comparison between our
result and the the fit to pionic atoms data is not direct. First of
all we consider only strong interactions and ignore effects due to
Coulomb forces, while these are crucial to bind the pion within the
pionic atoms on which the fit is based. They are discussed in
Ref.~\cite{KKW02}. Moreover not all renormalization effects are
included in our approach. For instance, we did not consider pion
rescattering on two correlated nucleons.  
It is known to provide a sizeable part of the
repulsive isospin-symmetric potential. Since it is not explicitly
introduced in the fits to data for the charge exchange potential,
 it has to be included as a
medium renormalization of $b_1$. This effect
 can be evaluated from the formulae given in
Ref.~\cite{ER72}:
\begin{equation}
\delta b_1^\mathrm{DS} = - (1+\frac{m_\pi}{M_N}) b_1 (b_0 +
b_1)\frac{3}{2\pi} p_F~,
\end{equation}
where $p_F$ is the Fermi momentum of the nucleons.
At normal density, it amounts to a decrease of $b_1$ by $\simeq 5\%$,
{\it i.e.}, it does not help to explain a larger enhancement. It would be
interesting to attempt to fit the pionic atom data with a more
moderate enhancement of the $b_1$ parameter. If the larger enhancement is
confirmed it would be an incentive to push the investigations further.
One might consider the influence of higher order effects in the
density, which could upset the nearly total cancellations present in
the isospin symmetric scattering length. For instance, the pionic piece
of the sigma commutator should be replaced by the fully dressed one in
which the pion line is dressed by nucleon-hole and $\Delta$-hole
excitations screened by short range correlations. At normal nuclear
density the effect leads to a moderate enhancement (of the order of 25\%) of
$\langle\phi^2\rangle$~\cite{CDDEM00}.  The many-body effects in the
other contributions should be similarly investigated, including those
concerning the phenomenological parameters $c'_2$ and $c'_3$, which
would require a model for these quantities. 
The question  of the in-medium
modification of the charge exchange $\pi N$ amplitude first raised by Weise is
indeed quite challenging on the theoretical side as well as on the experimental
one.  It is likely that more
physics remains to be understood in this problem.

\begin{acknowledgments} 
We thank A. Gal and W. Weise for stimulating discussions. One of us (M. O.)
acknowledges financial support from the Alexander von Humboldt-foundation as a
Feodor Lynen fellow.
\end{acknowledgments}



\begin{thebibliography}{99}
\bibitem{G02} H. Geissel et al., Phys. Rev. Lett. 88 (2002) 122301;
Nucl. Phys. A 663 (2000) 206.
\bibitem{KY01} P. Kienle and T. Yamazaki, Phys. Lett. B 514 (2001) 1.
\bibitem{W01} W. Weise, Nucl. Phys. A690 (2001) 98c.  
\bibitem{KW01} N. Kaiser and W. Weise, Phys.
Lett. B512 (2001) 283.
\bibitem{F02} E. Friedman, Phys. Lett. B524 (2002) 87;  Nucl. Phys A710 (2002) 117.
\bibitem{KKW02} E.E. Kolomeitsev, N. Kaiser, and W. Weise, nucl-th/0207090.
\bibitem{Y02} T. Yamazaki, Plenary talk given at PANIC02, Osaka (Japan) 28 Sept-
4 Oct 2002; to be published in Nucl. Phys. A.
\bibitem{CEG02} G. Chanfray, M. Ericson, and
P.A.M. Guichon, Phys. Rev C63 055202.
\bibitem{JCT92}I. Jameson, A.W. Thomas, and G. Chanfray, J. Physics G18 (1992) L159.
\bibitem{BMG92} M. Birse, J. Mc Govern, Phys. Lett. B292 (1992) 242
\bibitem{LTTW00} D.B. Leinweber, A.W. Thomas, K. Tsushima, and S.V. Wright, Phys.
Rev. D61 (2000) 074502.
\bibitem{CEW96} G. Chanfray, M.Ericson, and J. Wambach Phys.
Lett. B388 (1996) 673.  
\bibitem{CD99} G. Chanfray and D. Davesne, Nucl. Phys. A646 (1999)
125.
\bibitem{CEK94} G. Chanfray, M.Ericson, and M. Kirchbach, Modern Phys. Lett. A 9(1994) 279.
\bibitem{TW95} V. Thorsson and A. Wirzba, Nucl. Phys. A589 (1995) 633.
\bibitem{CDE98} G. Chanfray, J. Delorme, and M. Ericson, Nucl. Phys. 
A637 (1998) 421.
\bibitem{WOM02} A. Wirzba, J. Oller, and U.G. Meissner, Ann. Phys. 297 (2002) 27
\bibitem{S01} H.C. Schr\"oder et al., Eur. Phys. J C21 (2001) 473.
\bibitem{C02} G. Chanfray, Plenary talk given at PANIC02, Osaka (Japan) 28 Sept-
4 Oct 2002; to be published in Nucl. Phys. A.
\bibitem{CEO02} G. Chanfray, M. Ericson, and M. Oertel, work in preparation.
\bibitem{PJM01} T.S Park, H. Jung, and D.P. Min, nucl-th/0101064.
\bibitem{ER72} M. Ericson and M. Rho, Phys. Rep. 5C (1972) 59. 
\bibitem{CDDEM00} G.
Chanfray, D. Davesne, J. Delorme, M. Ericson, and J. Marteau,
Eur. Phys. J. A8 (2000) 283. 
\end{thebibliography}
\end{document}